\documentclass{article}
\usepackage{ijcai17}

\usepackage{times}
\usepackage{url}
\usepackage{amsmath}
\usepackage{graphicx}
\usepackage{graphics}
\usepackage{epsfig}
\usepackage{amsmath}
\usepackage{latexsym}
\usepackage{amsfonts}
\usepackage{amssymb}
\usepackage{paralist}
\usepackage{xspace}
\usepackage{setspace}
\usepackage{mathrsfs}
\usepackage{amssymb}
\usepackage{color}
\usepackage{algorithmic}
\usepackage[ruled]{algorithm}
\newtheorem{theorem}{Theorem}
\newtheorem{claim}{Claim}

\newtheorem{definition}{Definition}
\newtheorem{lemma}{Lemma}

\title{Towards Robust Monitoring of Stealthy Diffusion}
\author{Shaojie Tang\\
University of Texas at Dallas}

\begin{document}

\maketitle

\begin{abstract}
In this work, we introduce and study the \emph{$(\alpha, \beta)$-Monitoring} game on networks. Our game is composed of two parties an attacker and a defender. The attacker can launch an attack by distributing a limited  number of seeds (i.e., virus) to the network, and/or  manipulate the propagation probabilities on a limited number of edges.  Under our $(\alpha, \beta)$-Monitoring game, we say an attack is successful if and only if the following two conditions are satisfied: (1) the outbreak/propagation reaches $\alpha$ individuals, and (2) it has not been detected before reaching $\beta$ individuals. On the other end, the defender's ultimate goal is to deploy a set of monitors in the network that can minimize attacker's success ratio in the worst-case.  Our work is built upon recent work in security games, compared with stochastic guarantees, our adversarial setting leads to more robust solutions in practice.
\end{abstract}

\section{Introduction}
\label{sec:intro}

In this work, we use a network to capture the relationships and interactions within a group of individuals, e.g., it serves as the medium for the spread of infectious disease or computer worms. To model diffusion dynamics of an emerging infectious disease or computer worms within the underlying network, we adopt a simple and intuitive model called the \emph{Independent Cascade Model} (IC). This model was originally proposed in interacting particle systems \cite{durrett1988lecture,liggett2012interacting} from probability theory, and it has been widely used to capture the diffusion dynamics in many domains such as viral marketing. 

Due to incomplete knowledge of attackers' profile and plan, it is extremely difficult, if not impossible to predict their attacking strategy.  Notice that under IC model, the propagation probability on each edge plays an important role in predicting the diffusion process of an attack. Therefore, the attacker has strong incentive to manipulate this parameter by purposely affecting individuals' interaction preferences, and hence network structure.  In the example of spreading infectious disease, this could be done by creating certain events, such as a party, to change the frequency and distribution of individuals' interactions within a network. Take the spread of computer worms as another example, the attacker could  send some phishing email to the victim, indicating that successfully forwarding a virus-infected file to her friends could earn her some rewards. This somewhat changes the propagation probabilities between friends. In addition, the skillful attackers could adjust their strategy according to our current monitoring strategy. In this work, we assume that an attacker can launch an attack by (1) distributing a limited  number of seeds to the social network, and/or (2) manipulating the propagation probabilities on a limited number of edges. Take the spread of computer worms as one example.  Our model aims to capture the most powerful attacker, we can simply set that adjustable interval to be empty for those weaker attackers. All these have made it extremely challenging to design a robust monitoring strategy. On the defender's side, we are allowed to select a limited number of individuals as monitors, such that the attack can be detected as long as at least one of those monitors have been infected. We aim at designing the optimal selection of monitors to monitor the outbreak.

In this work, we introduce and study the \emph{$(\alpha, \beta)$-Monitoring} game.  We say a monitoring strategy succeeds if either of the following conditions are satisfied: (1) the outbreak fails to reach $\alpha$ individuals, or (2) the outbreak has been detected before reaching $\beta$ individuals. The ultimate goal is to find a monitoring strategy that can maximizes the expected gain in the worst-case.  
Our work is built upon recent work in security games, and we propose using  an algorithmic framework of double-oracle methods to solve our problem.

The main contributions are summarized as follows.

(1) A major contribution of this work is introducing a new problem, $(\alpha, \beta)$-outbreak monitoring,  that combines recent research in security
games and in influence maximization. The framework and method developed in this paper can be applied to a broad range of domains involving monitoring intentional outbreaks in a network such as cyber malware or rumor spread.


(2) Drawing from existing studies on security games, we propose using  an algorithmic framework of double-oracle methods. We propose algorithms for both the defender's and the attacker's oracle problems, which are used iteratively to provide pure-strategy best responses for both players. We present NP-hardness proofs for the defender oracle problem, and propose an approximation algorithm to solve it.

\section{Related Work}

Our work is closely related to competitive influence maximization and rumor blocking \cite{bharathi2007competitive,borodin2010threshold,he2012influence,tsai2012security,howard2010finding}.  However, our problem is different in three ways:
(1) Attacker's primary objective is beyond simply maximizing the number of infections. It is equally important to avoid being detected at early stage. (2) Defender's objective is  focusing on outbreak detection instead of blocking. (3) We consider a different, possibly the strongest, attacker model. First of all,  we assume that the attacker can select the seeds and manipulate the propagation probability on some edges. Most existing works assume that the attacker can only select a few seeds. Secondly, we treat the choice of attacker's strategy as adversarial rather than stochastic. Rather than one-player models, we are interested in generating equilibrium strategy.

Many existing work in security game focus on domains that were modeled as graphs \cite{basilico2011automated,jain2011double,halvorson2009multi,haghtalab2015monitoring}. Another closely related work is \cite{haghtalab2015monitoring}, their objective is to select a group of monitors that maximizes the detection probability of an outbreak before it reaches a target, however they focus on finding a pure strategy without featuring probabilistic components. In addition, they do not allow the attacker to manipulate the propagation probabilities. We are the first to study the adversarial outbreak monitoring problem that combines recent research in security games and in influence maximization.

\section{Diffusion Model and Problem Formulation}
\subsection{Network and Diffusion Model} In this work, we represent the  social network using a directed graph $G=(V, E)$. We use node set $V$ to represent the set of individuals and edge set $E$ to capture the relationships and interactions within a group of individuals, and $G$ serves as the medium for the spread of (infections) disease.  To model diffusion dynamics of an emerging disease within the underlying social network, we adopt a simple and intuitive model called the \emph{Independent Cascade Model}. The basic IC model can be roughly described as follows: The entire diffusion process starts with an initial set of infected nodes $S$, called \emph{seeds}, which could be intentionally selected by the attacker, and it unfolds in discrete steps in an randomized manner as follows. When node $v$ first becomes infected in step $t$, it has a \emph{single} chance to infect each currently un-infected neighbor $w$; it succeeds with a probability $p_{vw}$, a parameter called \emph{propagation probability}. If $v$ succeeds, then $w$ will become infected  in step $t + 1$, and the process is iterated till no more infections are possible. In the rest of this paper, let $\mathbf{p}=\{p_{e}|e\in E\}$ denote the propagation probability vector. One natural extension of this model is called \emph{repeated independent cascade model}, in which infected nodes remain active in all subsequent rounds, i.e., it attempts to infect all its uninfected neighbors in each round. All results derived in this paper can be extended to repeated independent model.

\subsection{($\alpha, \beta$)-Monitoring Game}
We formulate the monitoring problem as a leader-follower Stackelberg game. The defender (security agencies) acts first and the attacker (terrorists) observes the defender's strategy and then responds to it.

\textbf{ Attacker's Strategies.} We assume that the attacker can decide the distribution of seeds as well as adjust the propagation probability on some edges for its own benefit. Since it is not equally easy to adjust the propagation probabilities on different edges, we introduce the Adjustable Interval model  \cite{he2014stability} to capture to what extent this information can be manipulated on different edges. In particular, the propagation probability on each edge is associated an adjustable interval, e.g.,  we use interval $I_{vw}=[l_{vw}, r_{vw}]$ to present possible values of $p_{vw}$ that can be chosen by the attacker. The attacker's pure strategy is composed of two parts, the first part is to choose up to $c_1$ seeds, and in the second part, the attacker can manipulate the propagation probabilities on up to $c_2$ edges subject to each one's adjustable interval. Let  $\Theta$ denote the strategy space of the attacker, thus the attacker's pure strategy $\theta\in \Theta$ can be represented as
$\theta \triangleq \langle S_\theta, \mathbf{p}_\theta\rangle$ where $S_\theta$ is the set of seeds that have been selected under pure strategy $\theta$, and $\mathbf{p}_\theta$ is the manipulated propagation probability under pure strategy $\theta$. For ease of presentation, we call the attack launched under strategy $\theta$ as \emph{attack} $\theta$, and the resulting outbreak as \emph{outbreak} $\theta$. A mixed attacker strategy is a probability distribution over pure strategies, i.e., $\mathbf{y}=\{y_\theta | \theta \in \Theta\}$ with $y_\theta$ representing the probability that $\theta$ is selected.

\textbf{ Defender's Strategies.} The defender's pure strategy is to select up to $k$ monitoring nodes $M\in \mathcal{M}$, called monitors, where $\mathcal{M}$ is the defender's strategy space, composing of all subsets of nodes with size no more than $k$. We say an outbreak $\theta$ is detected by monitors $M$ if and only if at least one node from $M$ is infected by that outbreak. A mixed defender strategy is a probability distribution over pure strategies, i.e., $\mathbf{x}=\{x_M | M \in \mathcal{M}\}$ with $x_M$ representing the probability that $M$ is selected.

The attacker has two potentially conflicting objectives. On the one hand, the attacker wants to maximize its impact by causing a greater spread of infection, on the other hand, it is equally important for the attacker to minimize the likelihood to be detected before it reaches that goal. To capture this tradeoff, we introduce the $(\alpha, \beta$)-Monitoring Game, where $\alpha$ and  $\beta$ are two integers belonging to  $[1, |V|]$. We first define the $\alpha$-level outbreak as follows.
\begin{definition}
[$\alpha$-level Outbreak] An $\alpha$-level outbreak is any outbreak that infects more than $\alpha$ individuals.
\end{definition}
$\alpha$-level measures the maximum impact of an outbreak if it has not been detected or stopped, a larger $\alpha$ implies a greater spread of infection. We next formally define the utility model under $(\alpha, \beta$)-Monitoring Game. As in most existing works, we assume a zero-sum game.

\begin{definition}
[$(\alpha, \beta$)-Monitoring Game] Given both players' strategies $\mathbf{y}$ and $\mathbf{x}$
 \begin{itemize}
 \item The attacker's utility is defined as the probability that an $\alpha$-level outbreak has been successfully launched and it has not been detected before reaching $\beta$ individuals.
 \item The defender's utility is defined as the probability that either the outbreak fails to reach $\alpha$ individuals or it has been detected before reaching $\beta$ individuals.
 \end{itemize}
\end{definition}
 We next briefly discuss two special cases in order to give a better illustration of our model. The first special case is when $\alpha=1$ . Under this model, the attacker has minimum requirement on the impact of the attack, thus it poses even greater challenge on the defender's side since the only goal of the attacker is to avoid the detection. As discussed later, one immediate observation is that the attacker will never choose more than one seed since it can only increase the likelihood to be detected.  Another special case is when $\beta=1$. This setting again pose the greatest challenge on the defender's side. Regardless of the value of  $\alpha$, the only way to defeat the attacker is to detect the outbreak at the very beginning. Typically, we require that $\beta$ is smaller than $\alpha$ in order to compensate the reaction delays after the outbreak has been detected.

\textbf{Equilibrium.} Given both players' pure strategies $\theta$ and $M$, let $\rho_{\theta}(M)$ denote the utility function of the defender given both players' pure strategies $\theta$ and $M$, i.e., the probability that either the outbreak $\theta$ fails to reach $\alpha$ individuals or it has been detected by $M$ before reaching $\beta$ individuals.The objective is to find the mixed strategy $\mathbf{x}$ of the defender, corresponding to a Nash equilibrium of this monitoring game. Given the zero-sum assumption, i.e., the defender maximizes her minimum utility, the Stackelberg equilibrium is the same as the maximin equilibrium.  Thus the optimal mixed defender strategy $\mathbf{x}$ can be computed by solving the following linear program (LP).
\begin{center}
\enspace
\begin{minipage}[t]{0.45\textwidth}
\small
 \emph{ Maximize $\min_{\theta\in \Theta}\sum_{M \in \mathcal{M}} x_{M} \cdot \rho_{\theta}(M)$}\\
\textbf{subject to:}
\begin{align*}
\begin{cases}
\sum_{M \in \mathcal{M}} x_{M}=1\\
x_{M} \geq 0, \mbox{               } \forall M \in \mathcal{M}
\end{cases}
\end{align*}
\end{minipage}
\end{center}
\vspace{0.1in}
This LP has $|\mathcal{M}|$ variables which could easily be exponential in the number of nodes, and the attacker strategy space $\Theta$ could be infinite because each edge is associated with a \emph{continuous} adjustable interval. 
In the rest of this paper, we develop a double oracle based algorithm to solve the above LP effectively.

\section{Double-Oracle Approach}
In this section, we present KNIGHT, a double-oracle based algorithm for solving monitoring games. We also analyze the computational complexity of determining best responses for both the defender and the attacker. As subroutines of KNIGHT, we give the algorithms to compute the optimal or near-optimal responses for both players.
\subsection{Algorithm}
The double oracle framework \cite{mcmahan2003planning} first computes the equilibrium strategy for a smaller restricted game and then computes improving strategies for both players iteratively and eventually converges to a global equilibrium. Thus the key challenge reduces to computing the optimal or near-optimal responses for both players, which are called \emph{defender oracle} and \emph{attacker oracle}, respectively.

KNIGHT is listed in Algorithm \ref{alg:21}. $\mathcal{\overline{M}}$ is the set of defender strategies generated so far, and $\overline{\Theta}$ is the set of attacker strategies generated so far. \emph{CoreLP}($\overline{\mathcal{M}}, \overline{\Theta}$) computes an equilibrium of the two-player zero-sum game consisting of the restricted set of pure strategies $\overline{\mathcal{M}}$ and $\overline{\Theta}$, generated so far. The equilibrium over restricted strategy space can be solved efficiently as the strategy space is small. \emph{CoreLP} returns $(\mathbf{x}, \mathbf{y})$, which are the current equilibrium mixed strategies for the defender and the attacker over $\overline{\mathcal{M}}$ and $\overline{\Theta}$, respectively. The defender oracle (DO) generates a defender monitors $M$ that is a near best response for the defender against $\mathbf{y}$. Notice that $M$ is selected from all possible monitoring sets $\mathcal{M}$, which is not restricted to $\overline{M}$. The attacker oracle (AO) computes the best configuration $\theta$ against $\mathbf{x}$.

KNIGHT starts with a significantly small set of pure strategies for each player, and then expands these sets iteratively by applying the AO and DO to the current solution. Convergence is achieved when no improving strategy can be found for both players. Assume that both AO and BO can find the best response against the other player's current strategy, the solution obtained is optimal to the original problem. However, the above assumption does not hold in this work since it is NP-hard to find the best response for the defender, we develop an approximate DO that leads to an near optimal solution to the original problem.
\begin{algorithm}[hptb]
\caption{KNIGHT-Double Oracle for Robust Monitoring}
\label{alg:21}
\begin{algorithmic}[1]
\STATE Initialize $\mathcal{\overline{M}}$ by selecting an arbitrary set of monitors.
\STATE Initialize $\overline{\Theta}$ by selecting an arbitrary model.
\REPEAT
\STATE $(\mathbf{x}, \mathbf{y}) \leftarrow$ \emph{CoreLP}($\overline{\mathcal{M}}, \overline{\Theta}$);
\STATE $M \leftarrow$ DO $(\mathbf{y}_{\Theta})$;
\STATE $\mathcal{\overline{M}} = \mathcal{\overline{M}} \bigcup \{M\}$;
\STATE $\theta \leftarrow$ AO $(\mathbf{x}_{\mathcal{M}})$;
\STATE $\overline{\Theta}=\overline{\Theta} \bigcup \{\theta\}$;
\UNTIL convergence
\STATE Return $(\mathbf{x}, \mathbf{y})$
\end{algorithmic}
\end{algorithm}

\paragraph{CoreLP}
We first introduce the CoreLP, that is used to compute an equilibrium of the two-player zero-sum game consisting of the restricted set of pure strategies $\overline{\mathcal{M}}$ and $\overline{\Theta}$. The standard formulation for computing a maximin strategy for the defender in a two-player zero-sum game is listed as follows.
\begin{center}
\enspace
\begin{minipage}[t]{0.45\textwidth}
\small
\emph{CoreLP($\overline{\mathcal{M}}, \overline{\Theta}$):} \emph{ Maximize $_{\mathbf{x}}$  $U$}\\
\textbf{subject to:}
\begin{align*}
\begin{cases}
\sum_{M \in \mathcal{\overline{M}}} x_{M} \cdot \rho_{\theta}(M) \geq U,\mbox{               } \forall \theta\in \overline{\Theta}\\
\sum_{M \in \mathcal{\overline{M}}} x_{M}=1\\
x_{M} \geq 0, \mbox{               } \forall M \in \mathcal{\overline{M}}
\end{cases}
\end{align*}
\end{minipage}
\end{center}
\vspace{0.1in}
 The dual of CoreLP is listed as follows.
\begin{center}
\enspace
\begin{minipage}[t]{0.45\textwidth}
\small
\emph{Dual of CoreLP($\overline{\mathcal{M}}, \overline{\Theta}$):}  \emph{ Minimize $_{\mathbf{y}}$  $U$}\\
\textbf{subject to:}
\begin{align*}
\begin{cases}
\sum_{\theta\in \Theta} y_{\theta} \cdot \rho_{\theta}(M) \leq U, \forall M\in \mathcal{\overline{M}}\\
\sum_{\theta\in \Theta} y_{\theta}=1\\
y_{\theta} \geq 0, \forall \theta\in \overline{\Theta}
\end{cases}
\end{align*}
\end{minipage}
\end{center}
\vspace{0.1in}
 Recall that $\rho_{\theta}(M)$ is the probability that either the outbreak $\theta$ fails to reach $\alpha$ individuals or it has been detected by $M$ before reaching $\beta$ individuals. If we can estimate the value of $\rho_{\theta}(M)$ accurately, then CoreLP can be solved efficiently. Unfortunately, we next show that calculating the exact value of $\rho_{\theta}(M)$ is $\#P$-hard.
 \begin{lemma}
 Calculating the exact value of $\rho_{\theta}(M)$ is $\#P$-hard, even under the special case when $\alpha=1$ and $\beta=|V|$.
 \end{lemma}
\emph{Proof:} We prove this lemma by reducing the influence estimation problem to it. The input of a influence estimation problem is a social network $G=(V, E)$ and a propagation probability vector $\mathbf{p}$. Given any source node $s \in V$ and any target node $t\in E$, the objective is to calculate the probability that $t$ can be influenced or reached  by $s$ under IC model. We convert an arbitrary instance of the influence estimation problem to an instance of outbreak detection problem by constructing an attacker's strategy as $\theta \triangleq \langle S_\theta=\{s\}, \mathbf{p}_\theta= \mathbf{p} \rangle$, i.e., there is only one seed, and a defender's strategy as $M=\{t\}$, i.e., there is only one monitor. Since we assume that $\alpha=1$ and $\beta=|V|$, $\rho_{\theta}(M)$ is the probability that the outbreak starting from $s$ can be detected by $t$. Calculating $\rho_{\theta}(M)$ is equivalent to calculating the probability that $t$ can be influenced or reached  by $s$ in the given instance of the influence estimation problem.. It has been proved in \cite{chen2010scalable} that solving influence estimation problem is $\#P$-hard, thus calculating the exact value of $\rho_{\theta}(M)$ is $\#P$-hard. $\Box$

One standard approach to estimate $\rho_{\theta}(M)$ is using Monte Carlo simulation, however, running
such simulations are extremely time consuming. Instead, we can leverage a martingale approach developed in \cite{tang2015influence} to estimate $\rho_{\theta}(M)$ in near-linear time.

\subsection{Defender Oracle}
In this section, we describe the design of defender oracle. The defender oracle problem can be described as follows: generate the defender pure strategy $M$ (selecting a monitoring set) that maximizes the defender's expected utility against a given attacker mixed strategy $\mathbf{y}$ over $\overline{\Theta}$.
\begin{definition}[Influence Maximization Problem] The input of an influence maximization problem is a social network $G=(V, E)$ and a propagation probability vector $\mathbf{p}$. The objective is to select a number $k$ of nodes that maximizes the expected cascade size.
\end{definition}
\begin{lemma}
The Defender Oracle problem is NP-hard, even under the special case when $\alpha=1$ and $\beta=|V|$.
\end{lemma}
\emph{Proof:} Reduction from Influence Maximization problem to Defender Oracle: We convert an arbitrary instance of the influence maximization problem with social network $G$ and propagation probability vector $\mathbf{p}$ to an instance of the defender oracle problem by constructing the same social network with an reversed propagation probability vector  $\mathbf{p}'$, i.e., $p'_{uv}=p_{vu}$ $\forall u, v\in V$. We next construct the attacker strategy space $\overline{\Theta}$ and attacker mixed strategy $\mathbf{y}$ over $\overline{\Theta}$.
\[\overline{\Theta}=\{\langle \{v\}, \mathbf{p} \rangle|v\in V\}\mbox{ and }\mathbf{y}=\{y_\theta=1/|V||\theta\in \overline{\Theta}\}\] Under the above setting, the defender oracle problem is to find a monitoring set $M$ that maximizes $\sum_{\theta\in \overline{\Theta}}y_\theta \cdot \rho_{\theta}(M)=\frac{1}{|V|} \sum_{\theta\in \overline{\Theta}} \rho_{\theta}(M)$. Since we assume that $\alpha=1$ and $\beta=|V|$, $\rho_{\theta}(M)$ is the probability that the outbreak under $\theta$ has been detected by $M$. Because $\overline{\Theta}=\{\langle \{v\}, \mathbf{p} \rangle|v\in V\}$, each pure attacker strategy $\theta \in \overline{\Theta}$ contains a single seed, say $v(\theta)$. Together with the setting that $p'_{uv}=p_{vu}$ $\forall u, v\in V$, we have $\rho_{\theta}(M)$ is equivalent to the probability that $v(\theta)$ is influenced by $M$ in the given instance of the influence maximization problem. Therefore find a monitoring set $M$ that maximizes $\frac{1}{|V|} \sum_{\theta\in \overline{\Theta}} \rho_{\theta}(M)$ under $\mathbf{p}$ is equivalent to finding a set of $k$ nodes that maximizes the expected cascade size under $\mathbf{p}'$. $\Box$

\begin{lemma}
\label{lem:11}
Given a fixed attacker's pure strategy $\theta$, the defender's utility function $\rho_{\theta}(M)$ is submodular.
\end{lemma}
\emph{Proof:} We adopt a triggering set based approach introduced in \cite{kempe2003maximizing} to prove this lemma. Consider a point in the diffusion process under seed set $S_{\theta}$, we can view the outcome of this random event as being determined by flipping a coin of bias $p_{uv}$. In particular, for each edge $(u,v)$ in $G$, a coin of bias $p_{uv}$ is flipped at the very beginning of the diffusion process. The edges for which the coin flip indicated an successful activation are declared to be live in  $G$; the remaining edges are declared to be blocked in  $G$.
\begin{claim}
A node $u$ ends up getting infected if and only if there is a path from $S_{\theta}$ to $u$ consisting entirely of live edges.
\end{claim}
Each sample point, say $X$, in the probability space specifies one possible set of outcomes for all the coin flips on the edges.
Let
$
\rho_{\theta}^X(M)=
\begin{cases}
1, & \mbox{if the defender wins under $X$}\\
0, & \mbox{otherwise}
\end{cases}
$.
We next prove that $\rho_{\theta}^X(M)$ is submodular for every $X$. Consider two monitoring sets $A \subseteq B$, if the number of infected nodes under $X$ is smaller than $\alpha$, then $\rho_{\theta}^X(A)=\rho_{\theta}^X(B)=1$. This is because the attacker fails to reach $\alpha$ nodes, the defender always win. Thus $\forall v\in V: \rho_{\theta}^X(A\cup\{v\})-\rho_{\theta}^X(A)=\rho_{\theta}^X(B\cup \{v\})-\rho_{\theta}^X(B)=0$.
We next discuss the case when the number of infected nodes under $X$ is larger than $\alpha$.

 (1) If $A$ has detected the outbreak before it reaches $\beta$ nodes, i.e., some node in $A$ has been infected by $S_{\theta}$ before it reaches $\beta$ nodes, then $\rho_{\theta}^X(A)=\rho_{\theta}^X(B)=1$. This is because  $A \subseteq B$. Thus $\rho_{\theta}^X(A\cup\{v\})-\rho_{\theta}^X(A)=\rho_{\theta}^X(B\cup \{v\})-\rho_{\theta}^X(B)=0$.

(2) If $A$ has not detected the outbreak before it reaches $\beta$ nodes, but $B$ does, then  $\rho_{\theta}^X(A\cup\{v\})-\rho_{\theta}^X(A)\geq \rho_{\theta}^X(B\cup \{v\})-\rho_{\theta}^X(B)=0$.

(3) If neither $A$ nor $B$ have detected the outbreak before it reaches $\beta$ nodes, consider the newly added monitor $v$. If $v$ can not detect the outbreak before it reaches $\beta$ nodes either, then $\rho_{\theta}^X(A\cup\{v\})-\rho_{\theta}^X(A)= \rho_{\theta}^X(B\cup \{v\})-\rho_{\theta}^X(B)=0$. Otherwise, $\rho_{\theta}^X(A\cup\{v\})-\rho_{\theta}^X(A)= \rho_{\theta}^X(B\cup \{v\})-\rho_{\theta}^X(B)=1$.

Therefore $\forall v\in V: \rho_{\theta}^X(A\cup\{v\})-\rho_{\theta}^X(A) \geq \rho_{\theta}^X(B\cup \{v\})-\rho_{\theta}^X(B)$, it follows that $\rho_{\theta}^X(M)$ is submodular for every $X$. Now we are ready to prove that $\rho_\theta(M)$ is submodular. Since $\rho_\theta(M)$ can be represented as a linear combination of $\rho_{\theta}^X(M)$, i.e., $\rho_\theta(M)=\sum_{X}\Pr[X]\cdot \rho_{\theta}^X(M)$, then $\rho_\theta(M)$ is submodular because the linear combination of submodular functions is submodular. $\Box$

Lemma \ref{lem:11} together with the fact that  the linear combination of submodular functions is submodular imply the following lemma.
\begin{lemma}
\label{lem:22222}
Given an attacker's mixed strategy $\mathbf{y}$, the defender's utility function $\sum_{\theta\in \overline{\Theta}}y_{\theta}\cdot \rho_{\theta}(M)$ is submodular.
\end{lemma}
In the following, we propose a simple greedy algorithm (Algorithm \ref{alg:greedy-peak1}), which starts with the empty
set $M_0=\emptyset$, and in each iteration $i$, adds the node maximizing the marginal value to $M_{i-1}$ (ties broken arbitrarily):
\[M_i \Longleftarrow M_{i-1} \cup \{\arg\max_{v \in V\setminus M_{i-1}}\sum_{\theta\in \overline{\Theta}}y_{\theta}\cdot \rho_{\theta}(M_{i-1}\cup\{v\})\}\]

\begin{algorithm}[h]
{\small
\caption{Greedy-based Monitor Set Selection}
\label{alg:greedy-peak1}
\begin{algorithmic}[1]
\STATE $M_0=\emptyset; i=1;$
\WHILE{$i\leq k$}
\STATE $M_i \Longleftarrow M_{i-1} \cup \{\arg\max_{v \in V\setminus M_{i-1}}\sum_{\theta\in \overline{\Theta}}y_{\theta}\cdot \rho_{\theta}(M_{i-1}\cup\{v\})\}$
\STATE $i=i+1$;
\ENDWHILE
\RETURN $M$.
\end{algorithmic}
}
\end{algorithm}
A celebrated result by \cite{nemhauser1978analysis} proves that  when the objective function is submodular, the greedy algorithm provides a constant approximation to the optimal solution. Then together with Lemma \ref{lem:22222}, we have the following theorem.
\begin{theorem}
\label{lem:22}
Algorithm \ref{alg:greedy-peak1} provides a (1-1/e)-factor approximation for the Defender Oracle problem.
\end{theorem}
\begin{figure*}[hptb]
\hspace*{-0.7in}
\includegraphics[scale=0.21]{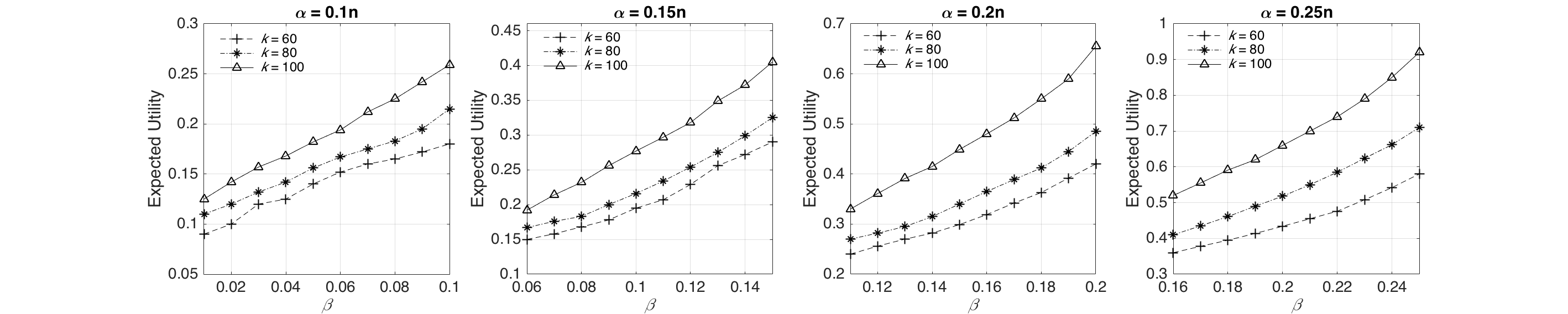}
\vspace{-0.3in}
\caption{Performance of the monitoring strategy measured by the expected utility of defender against varying $\beta$ under different settings of $\alpha$ and defender budget $k$.}
\label{fig:result}
\end{figure*}

\subsection{Attacker Oracle}
In this section, we describe the design of attacker oracle. The attacker oracle problem can be described as follows: generate the attacker pure strategy  $\theta = \langle S_\theta, \mathbf{p}_\theta\rangle$ (selecting a seed set $S_\theta$ and a propagation probability vector $\mathbf{p}_\theta$) that minimizes the defender's expected utility against a given defender mixed strategy $\mathbf{x}$ over $\overline{\mathcal{M}}$.

Since $c_1$ and $c_2$ are assumed to be some constant, we can enumerate all combinations of seeds and edges in time $O(|V|^{c_1} |E|^{c_2})$. However, the searching space $\Theta$ is still infinite because the adjustable interval of each edge is continuous. To tack this challenge, we next prove that our model can be reduced to an equivalent discrete space model.

\begin{lemma}
Under the PI model, given a fixed defender mixed strategy $\mathbf{x}$, the defender's expected utility is minimized by making each $p_{e}$ equal to $p_e-l_{e}$ or $p_e+r_{e}$.
\end{lemma}
\emph{Proof:} Pick any edge $\hat{e}$ and fix the propagation probabilities of the other edges. Consider any defender pure strategy $M$ and attacker pure strategy $\theta$, we define $\rho_\theta(M; \gamma)$ as the defender's expected utility   conditioned on the propagation probability of $\hat{e}$ is $\gamma$. Let  $\Pr[X|\gamma]$ denote the probability that $X$  happens when the propagation probability of $\hat{e}$ is $\gamma$. Then $\rho_\theta(M; \gamma)$ can be calculated as
$\rho_\theta(M; \gamma) = \sum_{X}\Pr[X|\gamma]\cdot \rho_{\theta}^X(M)$.

Inspired by \cite{He:2016:RIM:2939672.2939760}, let $\overline{X}$ (resp. $X\setminus \overline{X}$) denote the set of all blocked (resp. live) edges in $X$, then the probability  for obtaining $X$ conditioned on $\gamma$ can be calculated as:
{\small\[\Pr[X|\gamma] = \begin{cases} (1-\gamma) \prod_{e \in \overline{X}\setminus\{\hat{e}\}} (1-p_e) \prod_{e \in X\setminus \overline{X}} p_e \mbox{,  if $\hat{e}$ is blocked in $X$}\\
 \gamma \prod_{e \in \overline{X}} (1-p_e) \prod_{e \in (X\setminus \overline{X})\setminus \hat{e}} p_e\mbox{,             if $\hat{e}$ is live in $X$}\end{cases}\]}
In either case, $\rho_\theta(M; \gamma)$ is linear function of $\gamma$. Therefore the defender's expected utility under mixed strategy $\mathbf{x}$, i.e.,  $\sum_{M\in \overline{\mathcal{M}}} x_M\cdot \rho_\theta(M; \gamma)$, is also a linear function of $\gamma$, and thus minimized at one of the endpoints of the
interval. $\Box$

The above lemma shows that we only need to consider $O(|V|^{c_1} {(2|E|)}^{c_2})$ of possible attacker pure strategies.

\paragraph{What if $c_1$ is large?} We next discuss the hardness of the attacker oracle problem when $c_1$ is large.

\begin{lemma}
When $c_1$ is unbounded, the Attacker Oracle problem is NP-hard, even under the special case when $\beta=1$.
\end{lemma}
\emph{Proof:} This can be proved through reduction from set cover problem.
\begin{definition}[Set Cover Problem]
We can represent an instance of Set Cover
as a bipartite graph $G = (V_a, V_s, E)$, where $V_e$ is the set of elements, and $V_s$ is the set of subsets of $V_a$, and an edge $(u, v) \in E$ for $u \in V_a$ and $v \in V_s$ means $v$ contains $u$. The problem is to find a subset $S \subseteq V_s$ of minimize size such that all elements in $V_a$ are covered.
\end{definition}
We convert an arbitrary instance of the influence maximization problem with bipartite graph $G$ to an instance of the attacker oracle problem by constructing the social network $G$ with propagation probability vector defined as follows
\[\mathbf{p}=\{p_{vu}=1, p_{uv}=0| \forall (u,v) \in E\}\]
In the rest of the proof, we assume that $\beta=1$.  Notice that $\beta=1$ implies that there is no way for the defender to detect the outbreak on time, thus regardless of the deployment of the monitoring set, the attacker can always win as long as the outbreak reaches $\alpha$ nodes. Therefore, the attacker oracle problem is reduced to selecting a set of $m$ seeds that maximizes the probability of reaching $\alpha$ nodes.

We next prove that given a Set Cover instance, there exists a solution with size $k$ if and only if there exists a solution of attacker oracle instance with $m=k, \alpha=|V_a|+m, \beta=1$ such that at least $\alpha$ nodes can be reached with probability 1.  Assume the Set Cover instance has a solution with size $m$. Then every node in $V_a$ must be connected to some of those $m$ nodes in $V_s$. Consider the attacker oracle instance, the attacker can choose those $m$ nodes as seeds to reach at least $\alpha=|V_a|+m$ with probability $1$, i.e., the probability that the outbreak reaches $|V_a|+m$ is $1$. Conversely, if there exist $m$ seeds that  can infect $|V_a|+m$ nodes with probability 1, then it must be the case that all $m$ seeds are from $|V_s|$ and they can infect all nodes in $V_a$. It follows that those seeds can serve as a solution of the Set Cover instance.

To find the optimal solution of the Set Cover instance, we can solve the attacker oracle instance with $m=k, \alpha=|V_a|+m, \beta=1$ for $k=[1, |V_s|]$, and find the smallest $k$ that can reach $\alpha$ nodes with probability $1$, and corresponding seed set is an optimal solution of the Set Cover instance. $\Box$
\paragraph{Improved Results when $\alpha=1$}
We next discuss a special case of attacker oracle when $\alpha=1$. It is interesting to find that if there is no requirement on the scale of the outbreak, e.g, $\alpha=1$, then the attacker's objective is reduced to creating an outbreak that is least likely to be detected. Therefore, it is always preferable for the attacker to (1) keep the number of seeds small, and (2) reduce the propagation probabilities of certain edges to their minimum level. It follows that the attacker's best strategy is to select only one seed, e.g., $c_1=1$, and reduce the propagation probabilities of $c_2$ edges to their lower ends. Then the time complexity of enumerating all attacker pure strategies is improved to $O(|V| {|E|}^{c_2})$.
\subsection{Put It All Together}
Recall that if both DO and AO can be solved optimally, then the double oracle algorithm (Algorithm \ref{alg:21}) terminates with a solution which is guaranteed to be optimal. However, Theorem \ref{lem:22} shows that greedy is an $(1-1/e)$-approximate DO. We next extend the optimality results to approximate oracles in the following theorem, the proof is based on Theorem 2 in \cite{mcmahan2003planning2}.

\begin{theorem}
The Double Oracle approach returns a $(1-1/e)$-approximate mixed monitoring strategy.
\end{theorem}

\section{Empirical Evaluation}
To evaluate the performance of our monitoring strategy, we conduct extensive experiments on the \emph{Gnutella} datasets \cite{snapnets}. Gnutella is a peer-to-peer file sharing network. We simulate attacks and track the outbreaks on a snapshot of the network, with $8,114$ nodes and $26,013$ edges, where nodes represent hosts in the network topology and edges represent connections between the hosts. For independent cascade model, each edge is assigned a propagation probability $p_e$ randomly selected from $[0, 1]$. We set $l_e=r_e$ be a constant for each edge $e$ while satisfying the condition of $0\leq p_e-l_e\leq p_e+r_e\leq 1$. To obtain the value of $\rho_\theta(M)$, we leverage a martingale based approach developed in \cite{tang2015influence} to get an accurate estimation in near-linear time. In our experiments, the attacker can choose up to $c_1=30$ seeds and manipulate the propagation probability of up to $c_2=50$ edges. We vary the value of $\alpha$, $\beta$, and the budget of the defender $k$, to examine the effect of these parameters on the quality of our solutions. To obtain statistically sound results, we run the simulation $1,000$ rounds for each parameter setting, and the average value is reported in the following.

Figure \ref{fig:result} shows the comparison of the expected utility of the defender under different levels of an attack. Recall that an $\alpha$-level outbreak infects more than $\alpha$ individuals without being controlled. In a particular network, a larger $\alpha$ indicates a higher chance for the defender to win the game, since the offender has to infect more individuals to complete the attack. In our experiments, we vary the value of $\alpha$ from $0.1n$ to $0.25n$, where $n$ denotes the number of nodes in the network and $n=8,114$ for our dataset. As expected, we observe that the expected utility of the defender increases as $\alpha$ increases. We also examine the effect of $\beta$ on the expected utility. As shown in the figure, the expected utility is increasing with $\beta$. This is because, as $\beta$ increases, the defender can tolerant a larger spread of the attack in the network, leading to a higher chance of detecting the attack with a specific number of monitoring nodes. We also observe that as the budget of the defender $k$ increases from $60$ to $100$, the expected utility also increases, for the reason that with a higher budget, more monitoring nodes can be deployed to detect the attack, which leads to a higher chance for the defender to win the game.
\section{Conclusion}
In this work, we introduce and study the \emph{$(\alpha, \beta)$-Monitoring} game. Our goal is find a mixed monitoring strategy that maximizes the expected gain from the defender's side.  We study this problem under adversarial model without knowing attacker's strategy. To tackle this problem, we propose a novel monitoring strategy based on an algorithmic framework of double-oracle methods. We iteratively produce a single player's best-response to the opponent's strategy, and  add those strategies to the restricted game for the next iteration. If both player's best-responses can be computed in each iteration, then this approach terminates with a solution which is guaranteed to be optimal.  Our adversarial setting leads to a robust solution in practice.

\newpage
\bibliographystyle{named}
\bibliography{social-advertising-1}

\end{document}